# Proximity Scanning Transmission Electron Microscopy/Spectroscopy

Ing-Shouh Hwang

Institute of Physics, Academia Sinica, Nankang, Taipei, Taiwan

Email: ishwang@phys.sinica.edu.tw

**Abstract**

Here a new microscopic method is proposed to image and characterize very thin samples like few-layer materials, organic molecules, and nanostructures with nanometer or sub-nanometer resolution using electron beams of energies lower than 20 eV. The microscopic technique achieves high resolution through the proximity (or near-field) effect, as in scanning tunneling microscopy (STM), while it also allows detection of transmitted electrons for imaging and spectroscopy, as in scanning transmission electron microscopy (STEM). This proximity transmission electron microscopy (PSTEM) does not require any lens to focus the electron beam. It also allows detailed characterization of the interaction of low-energy electron with materials. PSTEM can operate in a way very similar to scanning tunneling microscopy, which provides high-resolution imaging of geometric and electronic structures of the sample surface. In addition, it allows imaging and characterization of the interior structures of the sample based on the detected transmission electron signals. PSTEM comprises a family of spectroscopies that address the transport and scattering of low-energy electrons in materials. Thus rich information can be extracted from the measurements. PSTEM offers a fundamentally new and powerful way to investigate thin materials. New analysis methods of thin materials and new physics may be uncovered by this method.

Scanning tunneling microscopy (STM) [1] has been a powerful technique to study topographic and electronic properties of sample surfaces with atomic resolution. Fig. 1a illustrates a schematic of STM. A metal tip is brought to a conductive sample surface to within ~ 1 nm. A bias is applied between the tip and the sample, resulting in a tunneling current between the tip and the sample surface. Depending on the polarity of the bias, the electron tunneling can be from the occupied states of the tip near Fermi level to the unoccupied states of the sample surface or from the occupied states of the sample surface to the tip. The electron energy of the tunneling electrons is typically smaller than 3 eV. Through raster scanning the tip over an area of the sample surface while maintaining a constant tunneling current, two-dimensional (2D) height profile of the surface can be obtained. Even though STM can reach atomic resolution on flat conducting surfaces, it can only probe the topmost atomic layer and is not sensitive to the structures underneath. In addition, STM often cannot resolve the atomic structures of molecules or clusters because STM is basically sensitive to the electronic structures rather than the atomic structures. A good example is STM imaging of DNA molecules. Numerous studies have shown that it is very difficult to use STM to resolve the base pairs of DNA.

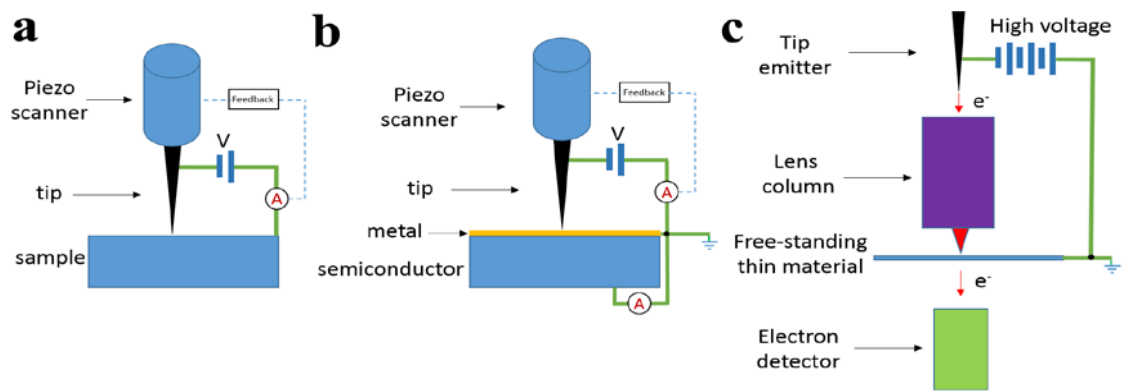

Fig. 1. Schematics of (a) STM, (b) BEEM, and (c) STEM.

After introduction of STM, a scheme of ballistic electron emission microscopy (BEEM) was introduced to measure the local Schottky barrier height [2], which is a three-terminal STM technique. When performing BEEM, the sample is a thin metal film (typically several nm thick) on top of a semiconductor, forming a Schottky diode. Electrons are injected from a negatively biased tip into the grounded metal film (Fig. 1b). A small fraction of these electrons will travel ballistically through the thin metal film to the metal-semiconductor interface where they will encounter a Schottky barrier. Those electrons with sufficient energy to surmount the Schottky barrier will be detected as the BEEM current. The atomic scale positioning capability of the STM tip gives BEEM nanometer spatial resolution. There are three electrodes with one to the thin metal film, one to the back side of the semiconductor, and one to the tip. The tunneling electrons injected from the tip are split into two components: one to the metal film and one reaching the underlying semiconductor.

Scanning transmission electron microscopy (STEM) has become a powerful tool in imaging and characterization of materials with a high spatial resolution. Typically electrons are field emitted from a sharp tip and the electron beam is focused by a lens column to a narrow spot, which is scanned over the sample in a raster using electric or magnetic fields (Fig. 1c). The transmitted electrons are used for image formation. Current STEM can reach atomic resolution at voltage higher than 20 kV thanks to the continuous advancement in aberration-correction techniques [3,4]. However, the focused spot size basically increases with decreasing electron energy and it becomes very difficult to focus the electron beam to atomic size below 10 kV. In addition to the high resolution imaging of the sample, STEM also allows several spectroscopic measurements, such as electron energy loss spectroscopy (EELS), to characterize the probe region.

STM and STEM have been considered as two very different techniques because the former is a proximity (or near-field) technique for surface characterization and the latter is a far-field technique to characterize the interior structures of materials. Here I would like to propose a new scheme, proximity STEM, which can be considered as a hybrid of STM, BEEM, and STEM. This method can operate like STM with a tip raster scanning over an area of the sample surface while the tunneling current between the tip and the sample is maintained constant through a feedback control. In addition, an electron detector behind the thin sample can detect the transmitted electrons (Fig. 2). Thus a 2D map of the transmission current can be acquired along with the typical STM image. The transmission image can be used to characterize thin materials, such as few-layer 2D conducting materials, and molecules or nano-objects adsorbed on or embedded inside the 2D materials. The tip is in proximity with the sample with a separation from tens of nanometer to sub-nanometer. No lens is used to focus the electron beam. The electron energy is typically below 20 eV.

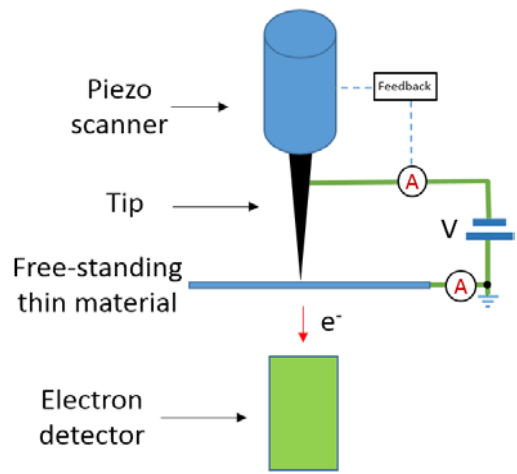

Fig. 2. Schematics of PSTEM.

There are three currents that can be measured. $I_e$ is the electron current emitted from the tip, $I_s$ is the current collected from the electrode connecting to the sample and $I_{tr}$ is the transmission current measured at the detector behind the sample.

Depending on the tip bias and the tip-sample separation, the measurements can be performed in either the tunneling regime or the field emission regime. In the tunneling regime, where the bias between the tip and the sample is typically below ~4 eV, the electrons from the occupied states of the tip tunnel to the unoccupied states of the samples. In the field emission regime, where the bias between the tip and the sample is typically above ~4 eV, the electrons from the occupied states of the tip tunnel to the vacuum states between the tip and the sample surface. A portion of the field emission electrons transports to the sample and can be collected, while another portion passes through the thin sample to the vacuum behind and gets collected by an electron detector. There are also electron reflected by the sample, but it is difficult to collect them due to the shielding of the tip shape. By scanning the tip laterally across the sample, one can measure the transmission current at different lateral positions of the sample. The local structure inside the sample can elastically and inelastically scatter the electron beam, giving rise to the contrast in the transmission current. For the transmission current, one can either measure the total transmission current or energy-resolved transmission current. Various spectroscopic techniques can be developed. For example, one can conduct the energy-resolved spectra of the transmission electrons at a specific sample position while keeping the tip-sample separation and the bias fixed. Also, one can ramp the tip bias to record the transmission currents. Furthermore, one can also apply the z-v measurements by measuring the tip-sample separation, z, while ramping the bias at a pre-set constant emission current. Generally, various STS measurements can be applied to PSTEM. Just now one can add the spectra of the transmission current. As in STEM, one can also obtain the EELS spectra by carrying out energy-resolved measurements of the transmission electrons, which provide information of the electronic and vibrational excitations of the local structures. The major difference between STEM and PSTEM is that the energy of the incident electron beam is higher than 20 kV for the former but lower than 20 eV for the latter. It would be very interesting to see the differences between the EELS spectra obtained at these two very different energy ranges.

    An advantage of PSTEM is that the electron inelastic mean free path (IMFP) can be several to tens of nanometers. Based on the universal curve of electrons (Fig. 3), the IMFP is below 1 nm between 20-500 eV, but it increases with decreasing energy below 20 eV. Thus a wider variety of thin specimens can be studied at a lower energy. For example, heterostructures of multilayer 2D materials usually have a thickness larger than 1 nm. In addition, one can use a suspended 2D material, such as monolayer graphene, as a support for biological molecules, such as double- or single-stranded DNA molecules. The diameter of a double-stranded DNA molecule is 2 nm, which is too thick to be penetrated by electrons with an energy in the range of 20-500 eV. This would not be a problem if the electron energy is reduced below ~12 eV. Over many years of effort, it remains difficult to resolve the individual base pairs of DNA with SPM techniques or electron microscopy. For the former technique, SPM measure the topographic and electronic structures of the surface, but structural information inside the molecules cannot be probed easily. For the latter, high-energy electron

beams often damage the molecules well before sufficient signal intensity is obtained for structural determination. If one places a DNA molecule on a free-standing graphene film. The transmission current in PSTEM may reveal structural and electronic information inside a material and thus different base pairs may exhibit different contrast. In addition, the defects or the local thickness of 2D materials may also be probed through analysis of the transmission electrons. Spectroscopy based on the field emission electrons from the tip and the transmitted electrons may provide rich information for characterization of a variety of properties inside the probed material.

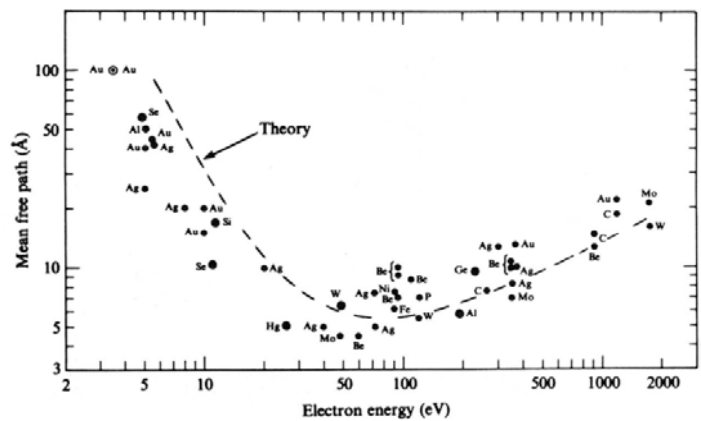

Fig. 3 Universal curve of electron mean free path [5].

The energy range below 20 eV was rarely explored in electron microscopy. It remains a question how low-energy electrons transport through and interact with materials. Even though IMFP has been measured at energies below 20 eV, but the measurements were conducted several decades ago. Furthermore, those measurements were indirect and had large variations. With PSTEM, now we can make more accurate measurements in a direct way and probe how IMFP varies with different materials and different local structures. There have been theoretical calculations of IMFP and the values vary with atoms. PSTEM will allow precise measurement of these parameters and interactions for comparison with theoretical calculations. An interesting question is how the heterostructures of multilayer 2D materials affect the transmission of low-energy electrons. Measurements of the transmission electrons may reveal whether there is any anomaly in the transport of low-energy electrons through the heterojuncions, such as reflection or other effects. New physics may be discovered from such measurements.

PSTEM also resembles the schematic of a lens-less electron projection microscopy (PPM) in some way. In PPM, the tip works as an electron point source and the magnified image of the sample is projected on the image plate [6-8]. A schematic of low-energy electron PPM is illustrated in Fig. 4. The magnification is equal to the ratio of the source-screen separation to the source-object separation. The PPM imaging is typically conducted at a tip-sample separation larger than 100 nm and the tip bias is in the range of 30-500 eV. In principle, PSTEM can be integrated with PPM if a 2D electron detector is used as the electron detector behind the sample. One can use PPM mode to image the sample when the tip is far away. During initial tip approaching, PPM imaging allows determination of the tip-sample separation based on the magnification of the images. PPM imaging also allow a larger view of the sample, which provides information for locating interesting areas for further tip approaching. When the tip is getting close to the sample, say within 1 μm, one can use the feedback system like the approaching mechanism in STM to do careful and fine approach to bring the tip to within a small separation from the sample, say 1-10 nm. Then one can switch to the PSTEM mode to operate. In sub-project 2, there is a lens-less system that can be used for such an integration test.

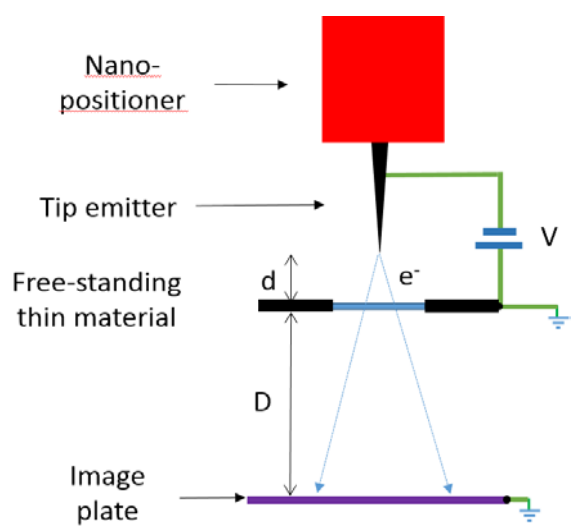

Fig. 4 Schematic of a PPM. The magnification of a PPM is (D+d)/d.

The tip in PSTEM is typically a sharp metal tip. A semiconducting or superconducting tip may also be used. One can also use a spin-polarized conductive tip to emit spin-polarized electron beam for probing magnetic materials adsorbed on or embedded in a suspended 2D materials.

For the sample preparation, one needs to prepare a suspended thin film, such as a 2D material. One way is to transfer the thin material to a sample holder containing holed regions to allow imaging in those regions. It can be a membrane (such as Si, $Si_3N_4$, $SiO_2$) fabricated with MEMS technology (Fig. 5a) or holey carbon film typically used as a TEM sample holder. The sample holder should be electrically conductive, which can be achieved either through doping or coating with a thin metal film. Transfer techniques for 2D materials to the sample holder have been reported by many research groups, because study of 2D materials has been an important subject in material research. For imaging individual organic molecules or nanostructures, one can use the few-layer structures as a support. A good support is clean few-layer graphene. Organic molecules or nanostructures can then be deposited onto the support for PSTEM study (Fig. 5b). The molecules or nanostructures can be deposited on either side of the 2D material. One can also prepare samples with molecules or nanostructures embedded inside the 2D materials, such as structures sandwiched between graphene layers.

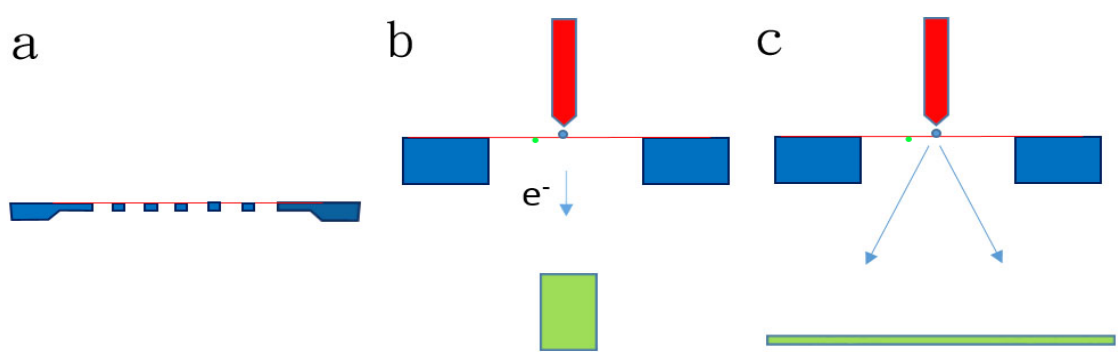

Fig.5 (a) Schematic of a sample holder containing holed regions (side view). A thin material (such as a 2D material) is placed on the holder. (b) Magnified view of a holed region with nano-objects adsorbed on either side of the free-standing thin material. A tip is placed in proximity to the sample (within several tens of nanometers but no physical contact) and an electron detector is placed behind the sample to collect the transmitted electron current. (c) A 2D electron detector is used to detect angle-resolved distribution of the transmitted electrons.

The electron detector can be a simple metal electrode, a Faraday cup, a channeltron, or a more sophisticated detector, such as an energy resolved or angle-resolved detector. Fig. 5b illustrates a scheme with an electron detector to measure the transmitted electron current. One can also carry out energy-resolved detection, such as the one using a retarding field spectrometer for electron detection [Fig. 1 in Ref. 9]. Fig. 5c illustrates another scheme to record the 2D intensity distribution of the transmitted electron beam. The detector is an image plate, such as a multi-channel plate (MCP). These angle- and energy-resolved spectroscopy may allow momentum- and energy-resolved measurements of the transmitted electrons, which contain structural and electronic information about the local area probed by the low-energy electron beam. In particular one can conduct EELS measurements, which allows determination of electronic and vibrational excitations, depending on the energy resolution of the detection.

Few-layer thin materials may exhibit significant vertical vibration near the center of the holed region. In addition, the tip-sample interactions, such as the van der Waals force and electrostatic force, may lead to local deformation of the material in response to the approaching tip. To enhance the mechanical stability of the thin materials, the diameter of each suspended region should be sufficiently small. One can use sample holders containing small holes. TEM holders containing membranes with periodic holes as small as 100 nm are already available in the market. The areas away from the central point of the suspended region should also have a higher stiffness for a better mechanical stability. In addition, one may cool down the sample to a certain cryogenic temperature to reduce thermal vibrations. Another advantage for cryogenic temperature operation is that the diffusion of small molecules (or clusters) on the thin film can be effectively suppressed, which allows study of small molecules or clusters.

It has been demonstrated that STM can image suspended graphene with atomic resolution [10,11]. Thus PSTEM should work as well. Furthermore, using STM-BEEM, Bannani et al. has shown that sub-molecular features can be resolved in the BEEM current images for molecules adsorbed on metal/semiconductor substrates [12]. The BEEM current measured the electrons that are transmitted through the molecules and the Schottky barrier between the thin metal film and the semiconductor. PSTEM also detects the transmitted electrons and should be able to achieve sub-nanometer resolution for molecules adsorbed on thin materials. In particular, the thickness of 2D materials (0.3-3 nm), such as graphene, is often much thinner than the thickness of the metal film in BEEM measurements (5-10 nm). Thus a better resolution can be expected for PSTEM images compared with BEEM images.

For 2D imaging of PSTEM, the tip can raster scan the sample using a scanner. As the operation in STM, it can be operated in a constant-current or constant-height modes. For the constant-current mode, one of the currents, $I_e$, $I_s$, or $I_{tr}$, is used as an input of a feedback control for maintaining a constant current during the scanning over the sample. In this way, the tip scan the sample surface with a nearly constant tip-sample separation and topographic images of the surface can be obtained. Meanwhile, other currents can be used for two-dimension maps. For the constant-height mode, either the feedback control is turned off or the feedback gain is reduced while the scan speed is increased, any one of the currents, $I_e$, $I_s$, or $I_{tr}$, can be used as the image signal. The constant-height mode is suitable for scanning a flat surface and the scan speed can be significantly increased. These 2D maps can be analyzed to understand the local structures and properties of the samples.

The electrical bias on the sample is grounded in Fig. 1c, but the sample can be a device with more than one electrodes. For example, there may be a voltage applied across the sample for heating or for actuation of the device. PSTEM can detect the response of the material when the voltage on the sample is changed. In addition, it is possible to provide external fields, such as electric, magnetic, or electromagnetic fields, during the PSTEM measurements. The responses of the thin materials under the external fields can then be investigated with PSTEM.

PSTEM allows development of a wide variety of spectroscopic methods (much more than STS). One can do analysis on either the emission current from the tip, $I_e$, the current collected from the sample, $I_s$, or the transmission current, $I_{tr}$. One can ramp the voltage of the tip and measure these currents. During the voltage ramping, one can either (i) fix the tip-sample separation and measure the $I_e$-V, $I_s$-V, and $I_{tr}$-V characteristics or (ii) fix $I_e$, $I_s$, or $I_{tr}$ and measure the V-Z curves. These measurements are similar to operation of typical STS, but now more currents are measured and analyzed. In addition, one can do energy-resolved measurement of the transmission current, which may provide EELS spectra. The transmission current provides crucial information about electronic, structural, and vibrational properties of the sample. Also, it allows characterization of interaction of low-energy electrons with local atomic structures of materials. Furthermore, one can use spin-polarized emitter or a superconducting tip to emit special electron beams for special characterization of samples.

Fig. 6 illustrates the signals generated when a low-energy electron beam interacts with a thin specimen. Analysis of the signals will allow development of new analysis methods and new physics may be uncovered by these methods. One may also detect photons emitted from the sample during PSTEM measurements. The photons allow analysis of electric and optical properties of the probed region. One can detect photons emitted on either the tip side or the transmission side. Due to the shielding effect of the tip shape, it would be easier to detect photons on the transmission side. In addition, one can apply the inverse photoemission spectroscopy (IPES) based on the PSTEM scheme to probe the unoccupied structure of the thin material with a high spatial resolution. Two modes of IPES can be used for this measurement. One is the isochromat mode, which scans the incident electron energy and keeps the detected photon energy constant. The other is the tunable photon energy mode, or spectrograph mode, which keeps the incident electron energy constant and measures the distribution of the detected photon energy.

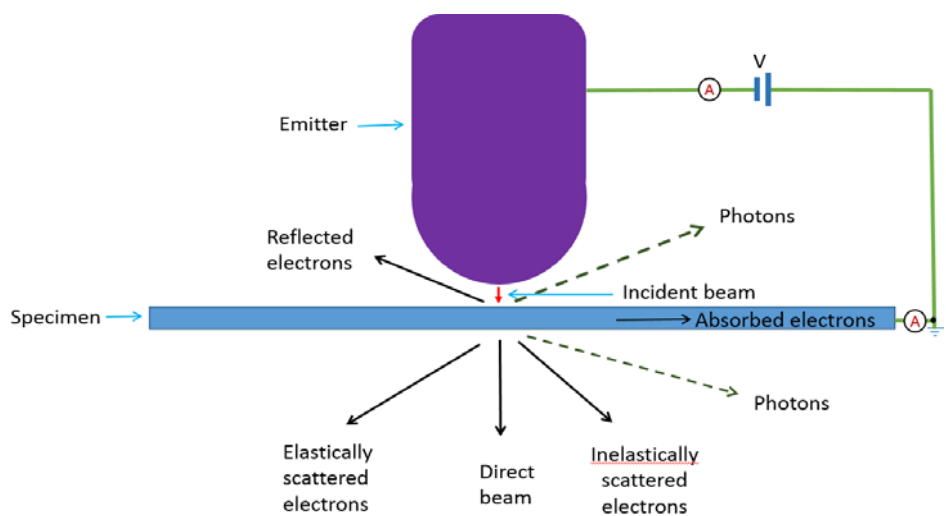

Fig. 6. Signals generated when a low-energy electron beam of electrons interacts with a thin specimen.